# Bubble dynamics in Liquid Hole Multipliers

A. Tesi[a,1], E. Segre[b], S. Leardini[c], A. Breskin[a], S. Kapishnikov[d],
L. Moleri[e], D. Vartsky[a] and S. Bressler[a]

[a]*Weizmann Institute of Science, Department of Particle Physics and Astrophysics, IL*
[b]*Weizmann Institute of Science, Physics Core Facilities, IL*
[c]*Instituto Galego de Física de Altas Enerxías, ES*
[d]*Weizmann Institute of Science, Department of Chemical Research Support, IL*
[e]*Technion - Israel Institute of Technology, Department of Physics, IL*

E-mail: andrea.tesi@weizmann.ac.il

ABSTRACT: In bubble-assisted Liquid Hole Multipliers (LHM), developed for noble-liquid radiation detectors, the stability of the bubble and the electro-mechanical properties of the liquid-to-gas interface play a dominant role in the detector performance. A model is proposed to evaluate the static equilibrium configurations of a bubble sustained underneath a perforated electrode immersed in a liquid. For the first time bubbles were optically observed in LAr; their properties were studied in contact with different material surfaces. This permitted investigating the bubble-electrodynamics via numerical simulations; it was shown that the electric field acts as an additional pressure term on the bubble meniscus. The predictions for the liquid-to-gas interface were successfully validated using X-ray micro-CT in water and in silicone oil at STP. The proposed model and the results of this study are an important milestone towards understanding and optimizing the parameters of LHM-based noble-liquid detectors.

# Contents

# Contents

# 1 Introduction

In the last years, advancements have been reached in the field of noble-liquid radiation detectors based on the bubble-assisted Liquid Hole-Multipliers (LHM) concept [1–5]. The original LHM concept [6], presented in the context of rare-event searches, was proposed to perform a combined detection of ionization electrons and primary scintillation photons generated by the interaction of radiation within a single-phase noble liquids. In its present configuration, an LHM-based detector consists of a CsI-coated perforated electrode (e.g. a Gas Electron Multiplier, GEM [7], or Thick GEM, THGEM [8]) immersed in the noble liquid with a stable gas bubble formed and trapped underneath. A typical LHM-detector layout is shown in Fig. 1. Radiation interaction in the liquid induces prompt scintillation-photon emission (S1) and ionization electrons. The latter, under appropriate electric fields, are focused into the holes and transferred from the liquid phase into the gas bubble; they induce electroluminescence (EL) under an electric field applied across the bubble, resulting in copious photon emission (S2) (e.g. $\sim$400 photons/e in Xe-LHM [9]). The S1 photons interacting with the CsI photocathode deposited on the electrode's top surface, result in photoelectrons emission; the latter are collected and transferred through the holes into the bubble, where similarly to the ionization electrons they induce S1' EL photons. The emitted light can be detected with an array of photo-sensors (PMTs, pixelated SiPMs etc.), to provide the event's energy and localization [4, 5]. For more details on the LHM concept and properties see [5, 9, 10].

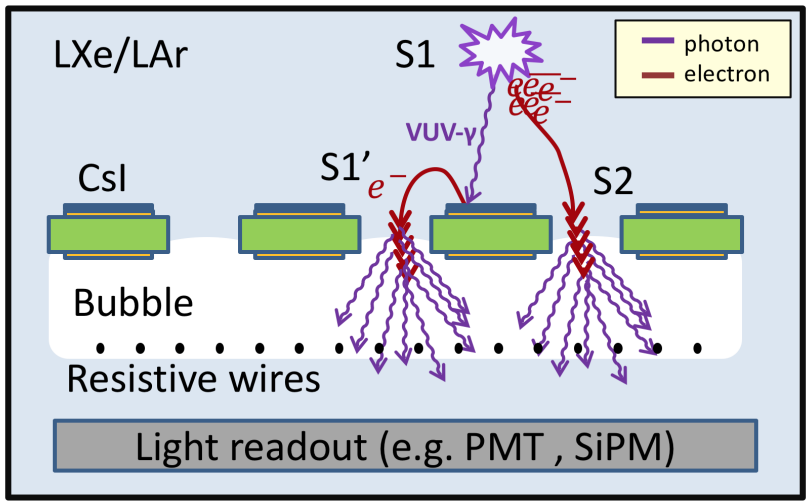

**Figure 1:** *LHM operation principle - A bubble is formed using a set of heating wires placed underneath a perforated electrode (GEM or THGEM). The electrode top is coated with a CsI photocathode. Ionization electrons focused into the holes, or S1 scintillation-induced photoelectrons emitted from CsI, create electroluminescence photons after crossing the liquid-gas interface into the bubble. A position-sensitive photon detector (e.g. SiPMs, PMTs), located underneath the wires, detects the resulting S2 and S1' photons. Typical dimensions for a THGEM electrode employed in our setup are: electrode thickness* 0.4 mm*, hole diameter* 0.3 mm *and hole pitch* 0.7 mm.

In this work we address questions related to the nature of the bubble trapped underneath the electrode. For example: which are the accessible equilibrium configurations for the bubble, what are the properties of the liquid-gas interface within the electrode holes (shape of the meniscus, protrusion) and how does the presence of an external electric field affect the bubble properties? These could affect the electron transfer efficiency from liquid into the bubble [9] and the EL processes occurring within the bubble - impacting thus on the final detector performance.

In Sec.2, we derived a theoretical model based on hydrostatic considerations for the equilibrium configurations of a bubble trapped underneath an ideal single-hole electrode, without the presence of electric fields. The model depends on the surface tension of the gaseous-liquid interface (LAr and GAr in our case) and the contact angle associated with the gaseous-liquid-solid junction (liquid and gaseous argon, and the solid-material substrate supporting the bubble). A measurement of the latter is described in Sec.3 [11].

The equilibrium configurations of the bubble, in the LHM configuration (of Fig.1), in the presence of an external electric field are studied in Sec.4 using accurate COMSOL™[12] simulations. A set of micro-CT scans of a perforated electrode immersed in silicone oil permitted to observe the shape of the bubble-to-gas interface and its variation under electric field applied across the electrode. These measurements are described in Sec.5. They permitted to validate the electrodynamical model and enable modeling the bubble properties in noble liquids, and to design optimal electrode and field configurations. This would allow optimizing both electron transfer into the bubble and the electroluminescence yield.

## 2 Equilibrium of the gas bubble under a THGEM-electrode hole: theory

We examine the properties of a gas bubble formed and sustained under a THGEM-electrode. These determine the shape of the resulting bubble, and the conditions for its eventual stability. In particular, it is important to know whether the electrode's hole contains gas or liquid and what is the shape of the interface separating the two phases. We can then describe the electric field distributions in gas and liquid, under potentials applied to the electrodes (shown in Fig.1), and discuss their role on the bubble stability and their effect on the detector performance. Thus, we consider the forces acting on the fluid, and examine the conditions that guarantee an equilibrium, taking into consideration the materials and geometrical parameters. In the following we assume that the fluid's gas and liquid phases are in static equilibrium. In an LHM-based detector, we might have to consider out-of-equilibrium conditions, responsible for the formation, the movement, the coalescence and the escape of such bubbles, as well as convection and pressure fluctuations within the cryostat. The treatment of all these dynamical effects is out of the scope of the present work.

We schematize the situation considering an ideal single-hole electrode with axial symmetry, with a standing bubble formed underneath. For a matter of graphic convenience, a bubble protruding from the top side of the electrode is illustrated in Fig.2 and discussed.

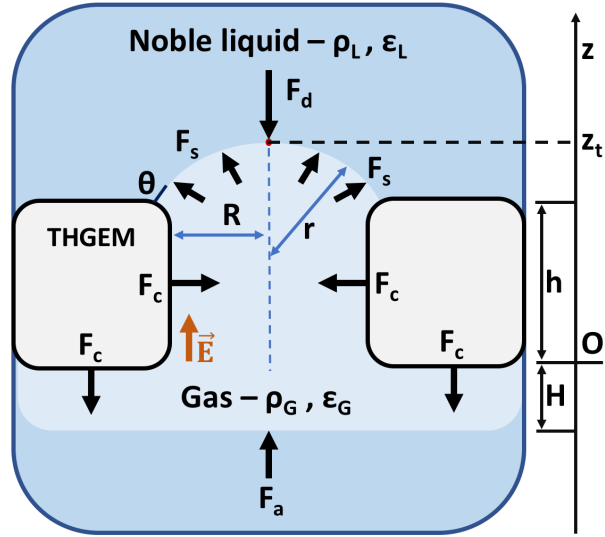

**Figure 2:** *Scheme of a bubble protruding from the top side of a bore including all the main forces responsible for the dynamical equilibrium between the liquid-gas phase.*

The bottom part of the bubble is considered flat and horizontal; its thickness $H$ is imposed by the amount of gas trapped and is considered fixed. The top part of the bubble is contained by the lower face of the electrode plate; part of it may penetrate the hole, with or without protrusion out of it - depending on the equilibrium configuration. The gas-to-liquid interface-meniscus (defined as "meniscus") is the only material surface which we consider as dynamically free within the fluid, i.e. it re-adjusts its shape in order to satisfy the prescribed balance of forces. At rest, all forces acting on the fluid in its bulk and at its boundaries, are in equilibrium. The relevant forces are (see Fig.2): a) the dielectrophoretic force $F_d$, due to the presence of a local electrical field $E$ and to different values of the electrical permittivity $\varepsilon$ in the liquid, $\varepsilon_L$, and in the gas, $\varepsilon_G$; b) the Archimedean lift $F_a$, due to the different densities of the gas and the liquid phases; c) the containment forces exerted by the solid support $F_c$, and d) the surface tension $F_s$, acting only at the interface between the two phases. Viscous forces, due to the differential motion of the fluid parcels, are absent in static conditions, and do not require consideration here.

The dielectrophoretic force acts throughout all the volume and it is generally expressed as the divergence of the Maxwell stress tensor, $F_d = \nabla \cdot \overline{\overline{T}}$, where $\overline{\overline{T}} = \varepsilon E : E - \frac{\varepsilon}{2} E^2 \overline{\overline{\delta}}$. We treat argon as an ideal dielectric, with neutral volume charge, and no charge accumulation at its interfaces between gas, liquid and solid boundaries. In this case, all terms related to the bulk drop out (see [13, 14]). The only remaining contributions arise from boundary terms at the material interfaces:

$$F_d = -\frac{\varepsilon}{2} E^2 \, n \qquad (2.1)$$

where $n$ is the local normal at the domain boundaries.

We first discuss the stability conditions for the upper bubble meniscus, in the absence of electrical field ($F_d = 0$). In first approximation we can derive an analytical expression for the meniscus position. The additional effect of the electrostatic field will be discussed later.

In purely hydrostatic conditions, the Archimedean lift $F_a$ is compensated everywhere in the bulk of the fluid by the gradient of the local pressure $p$. We can therefore simplify the treatment, imposing only the equilibrium of the boundary forces, and reason in terms of the pressure $p$ at the boundaries. On the solid immobile boundaries, a local constraint force balances the pressure, ensuring that the boundary remains fixed there, and we do not need to consider it explicitly. The force $F_s$ acting on the surface element of the gas-to-liquid interface is described as the Young-Laplace pressure

$$p_L = p_L n = \frac{2\gamma}{r} n \qquad (2.2)$$

where $\gamma$ is the surface tension between the two phases and $r$ the local radius of curvature of the meniscus. To describe both concave or convex menisci, we adopt the convention $r > 0$ if the center of curvature of the meniscus is above it, i.e. if the concavity is oriented upwards - and vice versa. With this convention we consider $p_L$ as positive if the force acts upwards. Referring to Fig. 3, all menisci depicted in green have $r < 0$, which implies downwards pressure forces and $p_L > 0$. A static meniscus implies

$$p_L = p_1 - p_2 \qquad (2.3)$$

where $p_1$ is the relative hydrostatic pressure exerted by the liquid out of the bubble and $p_2$ is the one exerted by the gas. At the bottom of the bubble, which we assumed flat, $p_{L,b} = 0$, and thus $p_{1,b} = p_{2,b}$. This in turn is equal to the hydrostatic pressure $\rho_l g H$ evaluated at the bottom side ($\rho_l$ being the density of the liquid and $g$ the acceleration of gravity). Inside the bubble, the pressure varies with height as $p(z) = p_{2,b} - \rho_g g(z + H)$. Here $\rho_g$ is the density of gas; it is treated in first approximation as incompressible. The small vertical dimensions of the system and the small variations of absolute pressure, when compared to the ambient pressure (the saturation pressure of argon at the operating temperature, as described below in Sec.3) justify this approximation. The $z$ axis is drawn upward, with its origin at the bottom of the THGEM-electrode plate. At the top of the upper meniscus ($z_t$) we have thus

$$p_{L,t} = -\rho_l g z_t - p(z_t) = (\rho_g - \rho_l) g (z_t + H) \qquad (2.4)$$

We interpret this as a condition imposing the radius of curvature of the meniscus as a function of its height $z_t$:

$$r = \frac{2\gamma}{(\rho_g - \rho_l) g (z_t + H)} \qquad (2.5)$$

We note that even for the simple case of an axisymmetric meniscus within a circular vertical tube (e.g. a classical capillary rise problem), no general analytical solution for the actual shape of the surface exists (a reference numerical solution is given by [15]). To further simplify, we neglect the piezometric height differences along the meniscus. This is justified because in our case they are small, of the order of the radius of the THGEM hole $R$, and typically smaller than the capillary length $\sqrt{\gamma/g(\rho_l - \rho_g)}$. We thus approximate menisci as circular caps of homogeneous radius $r$.

The location of the upper meniscus is determined by a further geometrical constraint, the attachment line on the solid substrate at the hole's rim. On this contact line the meniscus forms a prescribed contact angle $\theta$ (defined here in the liquid side, see Fig. 3) with the wall. This angle is an intrinsic feature of the chosen materials, i.e. the gas, the liquid and the substrate material. Some relevant measurements of $\theta$ are presented in Sec. 3. Fig. 3 schematically shows possible meniscus locations. To define the attachment point at the edge of the hole, the bore edges are considered smooth. The meniscus settles on a position which, additionally to the prescribed $\theta$, presents a downward concavity with the radius of curvature balancing the bubble pressure, and realizes a stable equilibrium with respect to movements normal to the interface center.

An approach for solving Eq. (2.4) is presented in Appendix A.1. In summary,

- for $\theta < 90°$, two equilibrium configurations are possible: the meniscus protruding from the top of the hole, or pinned at the bottom edge;

- for $\theta \geq 90°$, only a solution with the meniscus protruding from the top is acceptable.

A derivation of the solutions to the problem and the experimental material parameters of argon are reported in Appendix A.1. As results, it was found that the bubble thickness evaluates to $0\,\text{mm} < H < 7.7\,\text{mm}$ if the meniscus is pinned at the bottom edge and to $7.3\,\text{mm} < H < 11.1\,\text{mm}$ if the meniscus protrudes from the top of the hole.

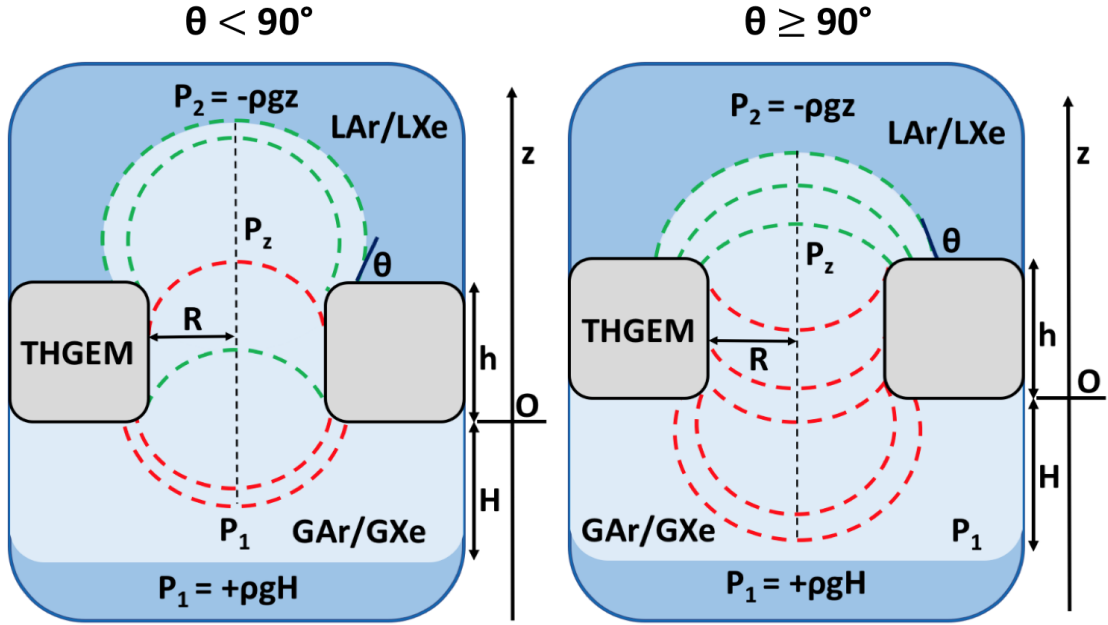

**Figure 3:** *Meniscus configurations for Left) contact angle $\theta < 90°$; Right) contact angle $\theta \geq 90°$. In green, several acceptable equilibrium configurations are shown. Dashed red lines represent non-equilibrium configurations.*

The electric field $E$ applied across the hole (see Fig.2) adds the term (2.1) to Eq.(2.3) and (2.4), transforming them into

$$p_L = p_1 - p_2 - \frac{1}{2}(\varepsilon_G E_1^2 - \varepsilon_L E_2^2) \qquad (2.6)$$

It thus requires a different value of the radius of curvature $r$ in order to satisfy Eq.2.6.

Formally, the electric force $F_d$, which also acts normally to the interface, can be assimilated to a local "electric" pressure force. In contrast to the piezometric term, the spatial dependence of the electric field is nontrivial for complex electrode geometries, and it is not as easy to define viable analytical approximations as for pure hydrostatics. We therefore resort to numeric computation, as described below in Sec.4.

## 3 Observation of captive bubbles in liquid Argon

The experimental characterizations of the bubble contact angles on different surfaces were conducted in a dedicated LAr cryostat, WISArD (Weizmann Institute Liquid Argon Detector). With respect to the previous experiments performed in [5], the cryostat system was modified to include 3 viewports pointing at the inner experimental assembly; they were oriented perpendicular, $+60°$ and $-60°$ with respect to the chamber axis. This cryostat was composed of an outer vacuum chamber (OVC) and by an inner one (IVC); the latter, of 100 mm in diameter and 150 mm height, was fillable with LAr (usually $\sim$ 250 ml). The two elements were separated by an interstitial gap kept under vacuum to provide stable thermal insulation to the IVC. The detector assembly was suspended from the topmost flange with an addional viewport.

The system was operated with LAr at the temperature of 90 K, with saturated vapor phase above, at a pressure $p_{sat} = 1.365$ Bar. The argon liquefaction process is described in details in [5]. In these conditions, the expected properties of argon are: liquid density $\rho_l = 1395.4$ kg/m$^3$, gas density $\rho_g = 1.784$ kg/m$^3$, surface tension $\gamma = 11.83$ mN/m, dielectric coefficient of the liquid phase $\varepsilon_l = 1.504$, dielectric coefficient of the gaseous phase $\varepsilon_g = 1.000516$ [16].

The experiments were prepared adopting the following methodology:

- Planar 1.5 mm thick, 7 mm in diameter slabs of different materials were prepared: Copper, Epoxy Resin (FR4), Teflon and Kapton. Surface finishing was realized using super-fine sanding (600-grits) in order to grant an high-quality polishing. Prior to their installation, the samples underwent an ultrasonic cleaning cycle in petroleum ether, acetone (whereas possible) and ethanol;

- A grid of resistive wires (Ni-Fe, 55 $\mu$m in diameter, 2 mm spacing) was fixed at 5 mm below the sample. The wires were powered to form bubbles in the noble liquid via thermal emission (Joule effect);

- Bubbles were observed with a CALTEX VIP-50-HD60 camera via the viewport orthogonal to the chamber axis; illumination was introduced through the other viewports (Fig. 4).

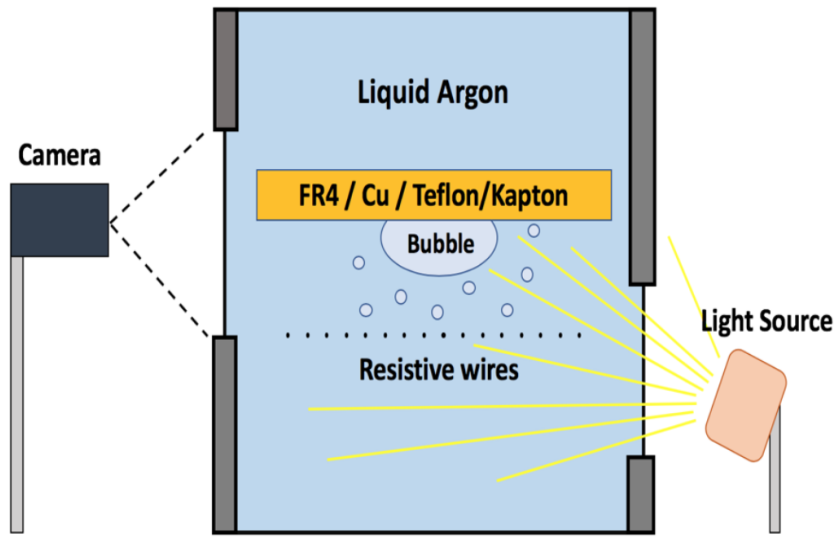

**Figure 4:** *Scheme of the setup used to characterize the contact angles of bubbles trapped under different materials in LAr.*

A standard technique to measure contact angles is the so-called "captive bubble method" [17], where the angles are reconstructed from the images of the bubbles trapped underneath a surface. In order to reconstruct the contact angle formed between the bubble and the chosen substrate, we recorded multiple sets of images (1000 frames/set); the camera operated in an automatic acquisition mode (1 frame/second). The images were processed using a dedicated Python script [18] (adjusting contrast and brightness, converting color into greyscale). Each set of images was subsequently manually analyzed and, selecting 200 frames containing distinct bubbles. Each sample was analyzed using the drop-shape analysis extension "DropSnake" [19] for the software ImageJ [20]. The shape reconstruction was done with an interactive spline method where the knots (see yellow dots in Fig. 5b, Fig. 5d) were placed by hand on the bubble interface [21]. This method offered us a rather straightforward and reliable way to reconstruct the bubble shape; it provided us with two, left and right, values of the contact angle for each frame. Each resulting statistical sample was thus composed of 400 angle values; these provided distributions, fitted to the obtained histograms by a dedicated Matlab script.

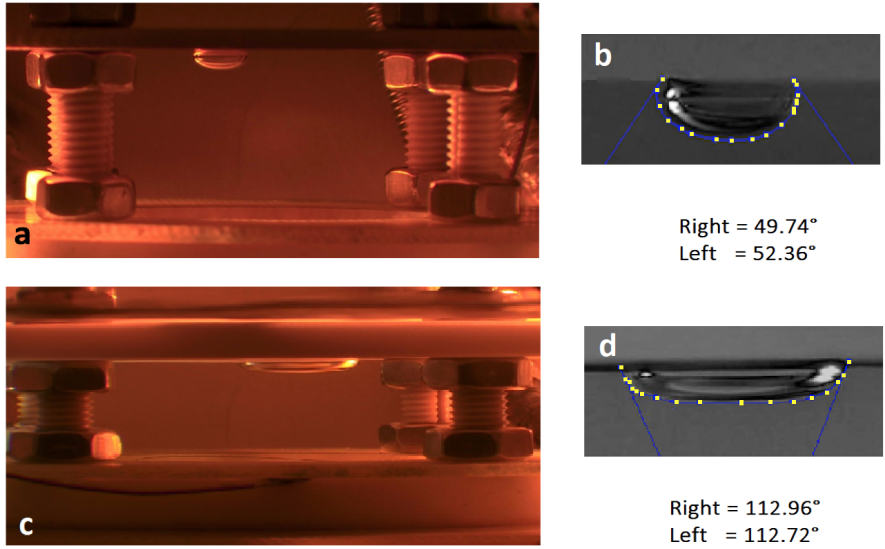

**Figure 5:** *(a) Argon bubble trapped underneath a Copper slab surface, immersed in liquid argon; (b) Bubble reconstruction from the image depicted in a), processed according to the methodology discussed in the text; both, left and right contact angles were recorded. (c) A bubble trapped underneath a Teflon slab immersed in LAr; (d) Example of bubble reconstruction of the image depicted in c).*

As discussed in Sec.2, according to the contact angle value, different equilibrium configurations are allowed, affecting the liquid-to-gas interface (meniscus location and shape). Conventionally, we will define the solid material to be:

- argon-phobic, if $\theta > 90°$, meaning that the local gaseous phase tends to adhere and spread over solid substrate (5d);

- argon-philic, if $\theta < 90°$, meaning that the liquid has the tendency to wet the solid material (5b).

The prefix "argon" in the definitions makes explicit reference to the liquid phase.

Our experiments were carried out with Copper, Epoxy resin (FR4), Teflon and Kapton. They revealed an argon-philic behaviour for bubbles sticking to FR4, Cu and Kapton (e.g. for Cu in Fig. 5 top); Teflon exhibited an argon-phobic tendency (e.g. Fig. 5 bottom). In the former case, the presence of small and escaping bubbles was favored while in the latter, the bubbles tended to coalesce into a large one and to spread over the substrate's surface.

Therefore, the measured contact angles are considered "dynamic" ones, since the bubbles did not remain static on a sufficiently long observation timescale. They rather represent an average between the "advancing" contact angle and the "receding" one. Precise values of "static" contact angles are therefore contained in an interval of values bounded by the advancing/receding contact-angle values [22].

An example shown in Fig. 6 depicts the Gaussian-fitted contact-angle histogram for bubbles sticking to a FR4 surface in LAr.

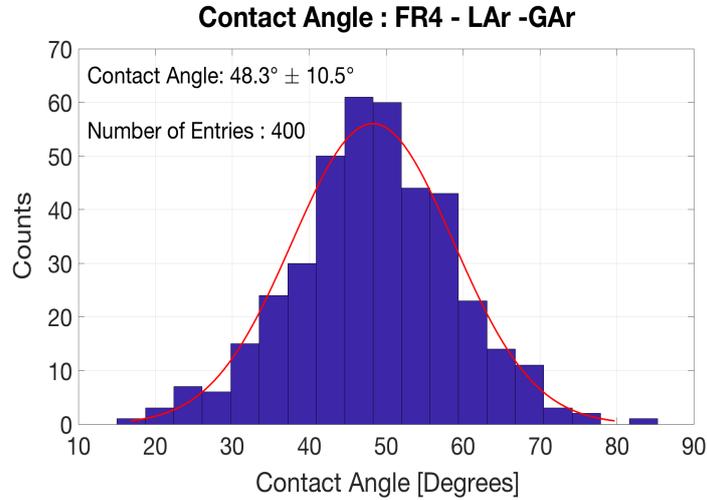

**Figure 6:** *Example of a fitted histogram of contact angles measured with argon bubbles sustained underneath a slab of FR4 surface, immersed in LAr.*

Table 1 summarizes the measured contact angles of bubbles sticking to Cu, FR4, Teflon and Kapton surfaces in LAr.

| Material | Measured $\theta$ | Behaviour |
|---|---|---|
| Copper | $45.5° \pm 8.5°$ | Argon-philic |
| FR4 | $48.3° \pm 10.5°$ | Argon-philic |
| Teflon | $108° \pm 6.5°$ | Argon-phobic |
| Kapton | $59.3° \pm 9.3°$ | Argon-philic |

**Table 1:** *Contact angles measured in LAr, for bubbles sticking to different substrate-material surfaces.*

Little can be said in quantitative terms about the effects of the environment surrounding the bubble (presence of fluid motions, convection, pressure gradients, spontaneous-bubbles background etc.). It was observed that the bubbles were "breathing" (cyclic deflation and swelling) [23] and occasionally escaping from the field-of-view during the measurements .

The dynamic contact angles (Table 1) measured here for the first time, are important parameters - implemented below in our numerical simulations.

## 4 Numerical simulations of the bubble electrodynamics

Determining the equilibrium conditions for the bubble in presence of an electrostatic field is crucial. The field, induced by the potentials applied to the various electrodes, depends on the medium (liquid, gas or substrate material), and on the geometrical layout. At the same time, the dielectric forces

– 10 –

on the fluid boundary affect the shape of the bubble, determining which parts of the domain are occupied by gas or liquid, changing the layout itself. In order to solve the problem, these two effects need to be coupled. Because of the non-trivial spatial dependence of the electric field, the problem is addressed computationally. Numerical simulations were carried out using the microfluidics module of the commercial package COMSOL™(v. 5.4). The numerical solution provides, in addition to the liquid-to-gas interface equilibrium shapes, full characterization of the electrical field in the vicinity of the hole rim. Full details about the protocol of the numerical simulations, and the methodology used to determine the stable bubble profile for given parameter values, are reported in Appendix A.2.

Fig. 7 depicts equilibrium positions of the meniscus in argon, obtained varying the electrode potentials, for menisci pinned at the top and bottom of the hole. Both sets use values $H$ of the bubble depth, and different applied-voltage range, for which pure hydrostatics guarantees stable solutions at $V_0 = 0$ V (see eqs. (A.5) and (A.2) of Appendix A.1).

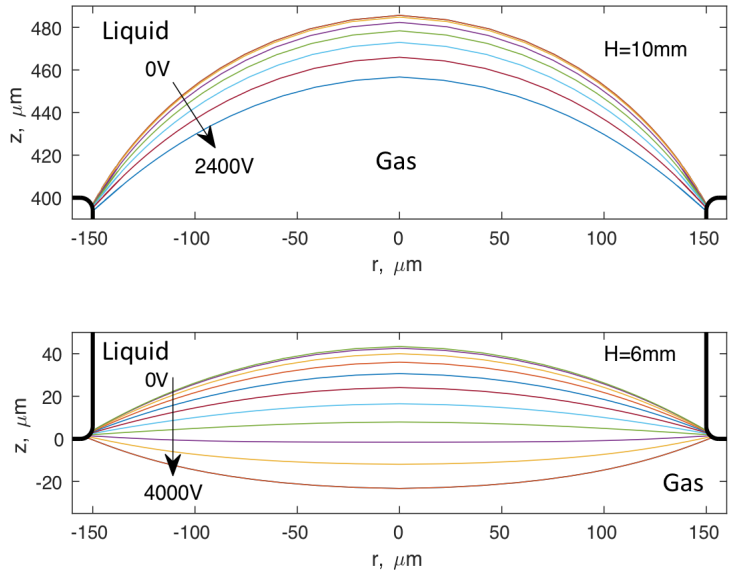

**Figure 7:** *Relaxed meniscus profiles in argon at (a) top and (b) bottom of the hole. Profiles are obtained at fixed bubble depths $H$, and varying $V_0$ values in steps of 400 V as indicated on the figure. $H$ and $V_0$ were chosen to satisfy equilibrium solutions.*

In the present geometry, both for meniscus locations at the top and at the bottom of the hole, the electric pressure (eq. (2.1)) always acts downwards from the upper liquid phase towards the gas one. Consequently, its effect is always that of increasing the value of the curvature $\kappa = 1/r$ (cfr. eq. (2.2)).

For increasing $V_0$ values, the concavity $\kappa$ is negative and is oriented downwards; in the case of the bottom meniscus, stable solutions with reversed and positive concavity exist for higher values of $V_0$.

Detailed analysis of the electric field, not reported here, shows local variations of the electric pressure ($P_E \sim E^2$), causing the formation of a non-uniform curvature. In a real experiment, deviations from the idealized geometrical shape induce discrepancies with respect to the computed profiles. For $\theta < 90°$, the meniscus can attach around the bottom edge of the hole to different points than at the top. This implies a larger range of admissible curvatures satisfying the contact angle constraint and means that it is physically possible, tuning parameters like $H$ and $V_0$, to cause a larger variation of the meniscus concavity if the same is attached to the bottom of the hole rather than at the top. On the lower face, both convex and concave menisci are allowed (cfr. Fig. 12a and 12c in the Appendix). This is reflected in the larger range of applicable voltages to maintain equilibrium at the bottom, albeit for different values of the gas layer thickness $H$ (keeping all other material parameters equal).

We remark that the method used here is essentially one that searches for the relaxation to a static bubble equilibrium, and serves only to determine the parameter ranges for which the one or the other meniscus solution is stable, and the resulting actual bubble shape. The numerical procedure adopted here does not reveal which of the two positions of the meniscus is more likely to appear when the bubble is formed, and does not guarantee that a bubble destabilized by a large perturbation will return to the same equilibrium shape. It is for example possible that a bubble collapses to a given different position once some amount of gas is added at the bottom or escapes through the hole. Such effects can only be assessed by full fluid dynamics simulations, which are beyond the scope of this work.

## 5 Experimental validations by X-ray micro-CT

The validation of the results derived from the model presented in Sec.2, requires a direct observation of the bubble interface location and shape. While not simple in a noble liquid, a way was found to observe the interface in a perforated electrode immersed in water or in silicone oil, at room temperature – applying X-ray microtomography ($\mu$-CT) [24, 25]. This technique, widely employed in the fields of biomedicine, materials research and in industry [26–28], offers the possibility to reconstruct with micrometric precision 3D tomographic images of an object using X-rays; these represent a 3D distribution of the mass attenuation coefficients within the object.

A successful validation in these materials of the bubble hydrostatics and electrodynamic models, described above, is expected to predict the behavior of bubbles in the noble liquid.

### 5.1 Validation of the bubble hydrostatics in water and in silicone oil

The perforated-electrode sample used in the hydrostatic experiments was composed by a 0.4 mm thick FR4-made bare (no Cu) THGEM-like disc of 13.9 mm diameter with 0.5 mm diameter holes , spaced by 1 mm. The disc was installed in a dedicated Plexiglas container, with a side aperture for inflating a gas bubble underneath. Two liquids were selected for the investigations: water (with 5% KI added to improve the contrast) and silicone oil (Dow Corning diffusion pump oil - 704); the latter was chosen because of its electrical insulating properties (for further experiments under electric fields). The assembly is depicted in Fig. 8. An air-bubble was injected underneath the electrode immersed in the liquid. Then, the container was introduced into a ZEISS Xradia 520 Versa micro-CT scanner (up to 0.7$\mu$m spatial resolution, X-ray source: 30-160kV). The tomographic sets of X-ray transmission images were acquired in 1201 consecutive steps over an angular range of 360 degrees at the X-ray source voltage of 110kV and the current of 91 $\mu$A. Each transmission image with the pixel size of 2.1$\mu$m was acquired over 22 s, whereas an entire tomographic set was collected over ca. 8 h.

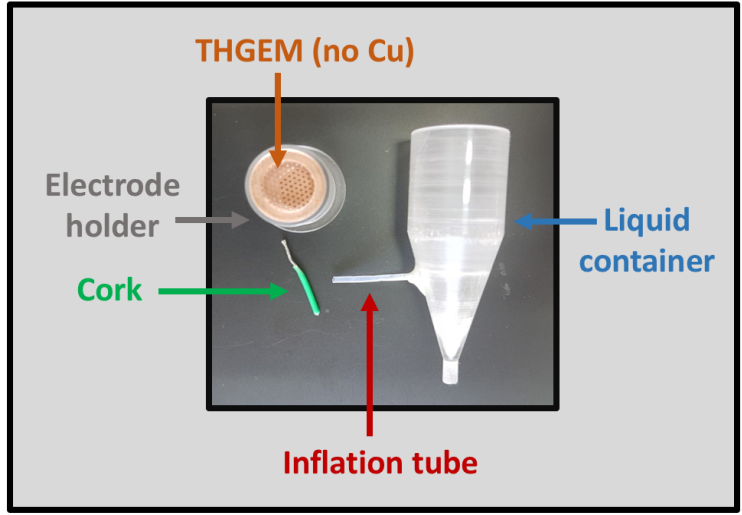

**Figure 8:** *Bare THGEM sample installed into a dedicated Plexiglass electrode holder.*

Reconstructions of the bubble-interface protrusions into the holes obtained in the two investigated liquids are illustrated in Fig. 9. The bubble interface penetrates into the 0.5mm diameter holes by $\approx$ 16.5 $\mu$m and $\approx$ 37 $\mu$m in water and oil, respectively. These values are in good agreement with the respective model predictions of 16 $\mu$m and 35 $\mu$m. The thickness of the bubble in these experiments was $\approx$ 4 mm. Estimations of the expected protrusions were computed for both liquids using the values of the contact angles measured on a FR4-substrate surface with a sessile drop method ([29, 30]). From the contact angle measurements, it was found: $\sim$39° and $\sim$24° for water and silicone oil, respectively. Possible causes for the very small deviations ($\approx$ 6%) in the interface protrusion from our model prediction, might be related to uncertainties in the measured contact angles or other experimental systematic errors.

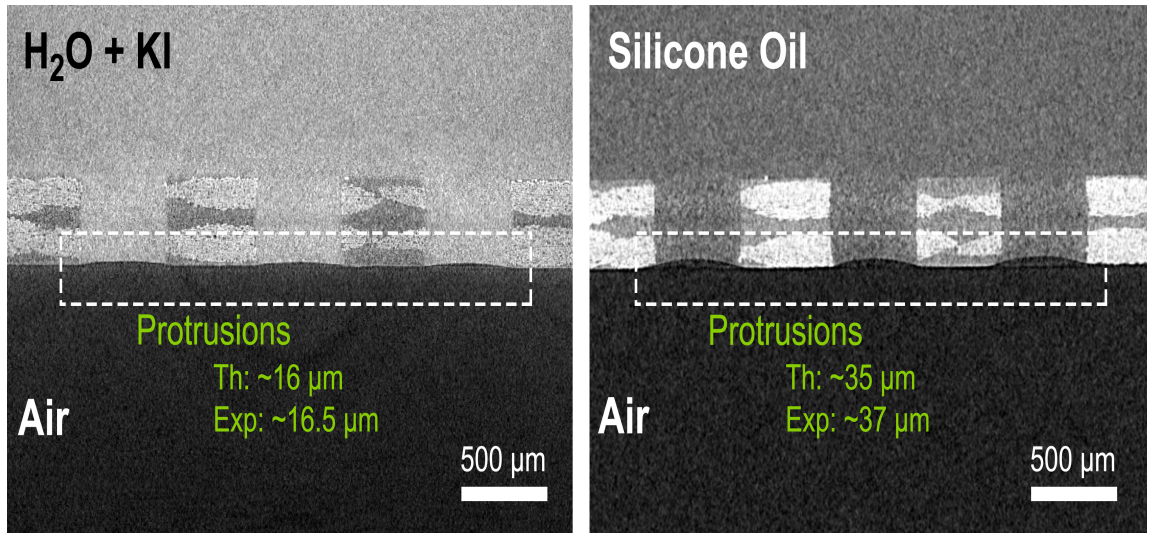

**Figure 9:** μ-CT cross-section images of air bubbles trapped underneath a bare perforated electrode immersed in Left) H$_2$O+5%KI; Right) Silicone oil. The theoretical and experimental values for the bubble protrusion into the 0.5mm diameter holes are displayed. Note the laminated structure of the FR4 epoxy-resin plate.

## 5.2 Validation of the bubble electrodynamics in silicone oil

Following the hypothesis that the electric field at the interface should have an impact on its shape and the model-simulation results, led our experimental efforts to validate the bubble electrodynamics. Experiments were carried out in silicone oil, using the same container. To avoid X-ray beam hardening artifacts in the μ-CT reconstruction process, metal coating of the perforated electrode had to be avoided. Instead, the two surfaces were evaporated with thin (10nm) graphite films prior to hole drilling. In addition, the 1mm-thick FR4 perforated-electrode disc was designed with a 0.1 narrow Cu HV-contact ring.

After graphite-film evaporation, holes of 0.8 mm diameter were drilled on the disc. The experimental μ-CT procedure is similar to the one described above. The voltage was supplied to the electrode by a CAEN N8315 NIM Power Supply. Tomographic scans were run at two different voltage configurations: ΔV= 0 kV and 3kV (E = 0-30kV/cm over the 1mm thick disc). The tomographic scans were of 8h duration and of an intrinsic resolution of 2.5μm.

It should be added that the contact angle of an air bubble trapped in silicon oil underneath a graphite surface was measured by x-ray transmission (see Fig. 10) in the microtomography system; it is an important parameter for the model simulations.

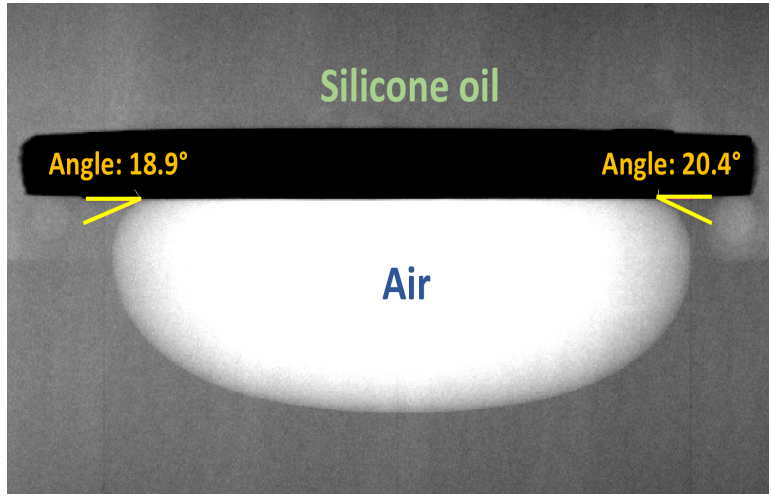

**Figure 10:** *μ-CT reconstruction of an air bubble trapped underneath a graphite-coated disc immersed in silicone oil. The average measured contact angle equals to 19.6°.*

The measured average value for the contact angle of the air bubble sustained below a graphite-coated disc immersed in silicone oil is 19.6° and the measured bubble thickness is ≈ 3.4 mm.

In order to implement our simulations, the following parameters were taken into account: the measured contact angle, the dielectric constant of air $\varepsilon_{air} = 1$, of silicone oil $\varepsilon_{SO} = 2.8$ ($\Delta\varepsilon_r = \varepsilon_{air} - \varepsilon_{SO} = 1.8$) and the surface tension of silicone oil $\gamma_{SO} = 37.3$ mN/m, as reported in [31], [32].

A direct comparison of the reconstructed protrusions via $\mu$-CT with the simulations was not possible because of the presence of surface imperfections stemming from the drilling process. Instead, we opted for the comparison of the reconstructed protrusion variations measured at the center with the electric field assuming ideal edges. The reconstructed images for the applied voltage $\Delta V$=0kV and 3kV are shown in Fig. 11.

The difference between the expected protrusion variation within the 0-3 kV range in simulations, ∼28 $\mu$m, and the measured one, ∼34 $\mu$m, is in agreement within ∼16% error. It is possible to observe that not only the interface re-adjusts itself to satisfy the new equilibrium set by the presence of the electric field, but also that the preferential configuration corresponds to a bubble protrusion into the hole from the bottom of the electrode. No bubble protrusion was observed in this work from the top side of the electrode.

In Fig. 11, the reconstructed images for the applied voltage $\Delta V = 0$kV and 3kV are shown.

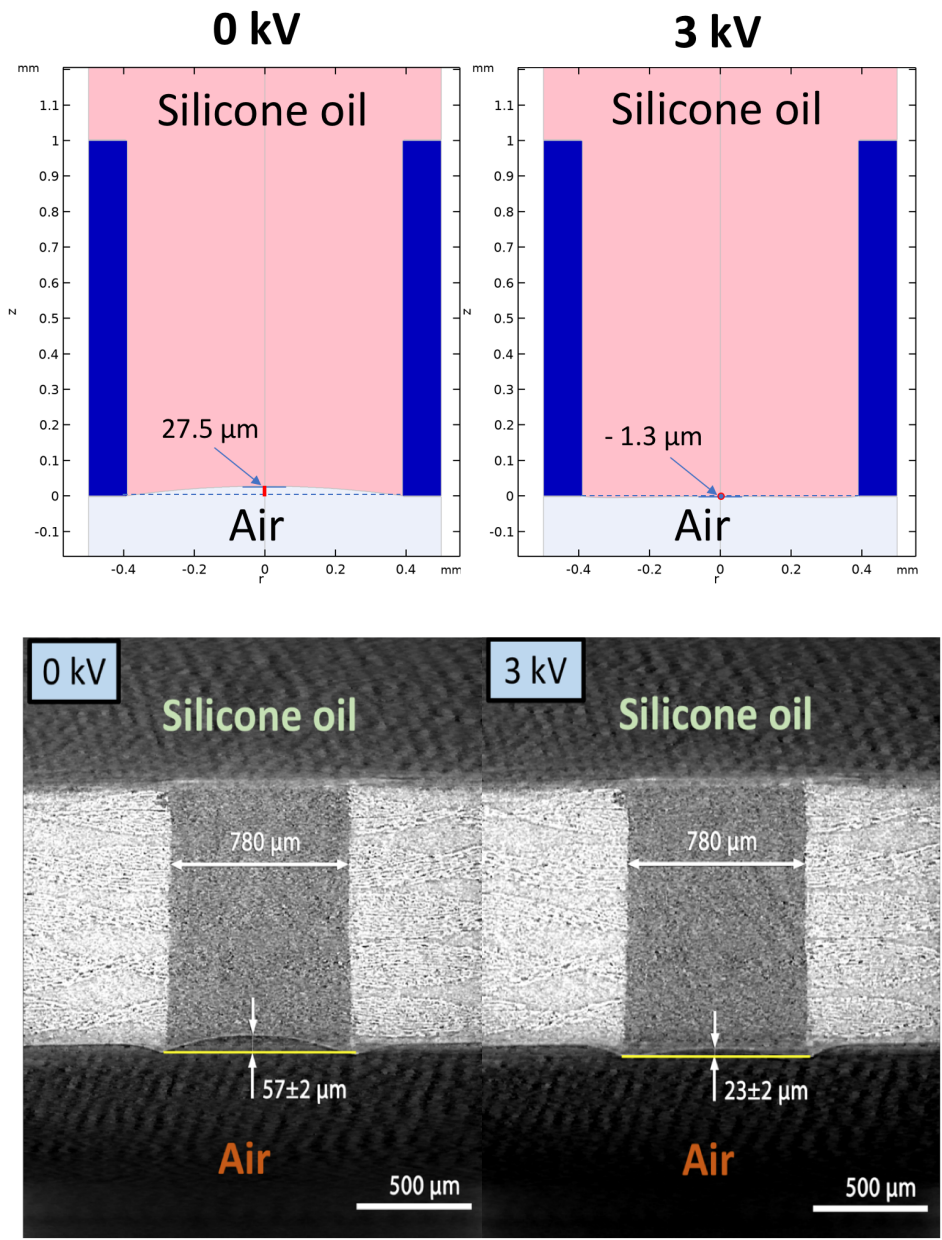

**Figure 11:** *Top) Results of the simulation of an air bubble trapped under a graphite-coated THGEM hole immersed in silicone oil for two different voltage configurations; Left: 0 kV; Right: 3 kV.* **Bottom)** *Reconstructions of an air bubble trapped under a graphite-coated THGEM immersed in silicone oil for two different voltage configurations; 2.5µm resolution scans. Estimations of the experimental bubble protrusion variations are displayed on each panel. Left: 0 kV; Right: 3 kV.*

## 6  Summary and discussion

In this work, we aimed at evaluating the nature of the interface of a gas bubble trapped in liquid under a perforated electrode. The subject is of high relevance for understanding the physical processes in bubble-assisted Liquid Hole Multiplier (LHM) noble-liquid radiation detectors. Starting from the hydrostatic equations describing a bubble trapped in liquid argon underneath an electrode, the possible bubble equilibrium configurations were computed and the distinct behaviors (argon-philic/-phobic) dictated by the value of the contact angle were discussed. For the first time, bubbles in liquid argon were optically observed and their dynamical contact angles were measured in liquid argon on potential materials forming LHM electrodes: Cu, FR4, Teflon and Kapton. The experimental results opened the way to investigate with numerical simulations the effects of external electric fields on the bubble equilibrium configurations and on the meniscus-like liquid-to-gas interface in the presence of holes. As a result, it was clarified that the electric fields at the interface vicinity contribute to the equilibrium balance as an additional pressure term. This translates into a downwards thrust and into deformations of the bubble interface according to the intensity of the applied field. The results of the model-simulated bubble hydrostatics and electrodynamics in room-temperature water and silicone oil, were validated by $\mu$-CT imaging experiments. It is of interest to report here that in all the experiments performed at room temperature no stable bubble protruding from the top side of the electrode was observed for $\theta < 90°$. The presence of a liquid-to-gas interface located at the bottom of the hole was also outlined in [9] in the context of LXe-LHM. Therefore it would be reasonable to assume that also in LAr, for $\theta < 90°$, the expected equilibrium configuration corresponds to a bubble protruding from the bottom side of the electrode's hole. Using $\mu$CT it was shown that it is possible to reconstruct the interface of an air bubble trapped underneath an electrode immersed in water or in silicone oil within 6% agreement with the theoretical predictions. A validation of our bubble electrodynamics model was achieved in silicone oil, under different electric-field conditions; the observed interface-protrusion variations with the field were in 16% agreement with the simulations. The successful validation of our model in room-temperature liquids, should lay the groundwork for inferring the bubble-interface configuration in noble liquids. Further model-simulations studies are expected to shed more light on the interface modifications induced by shaping the electric fields at the holes' vicinity. These could serve as a solid base for the enhancement of the electron transfer efficiency from a noble liquid to the bubble in LHM detectors.

An optimal field (and interface) shaping could also play a role in enhancing their electroluminescence (EL) photon yield. While currently, most of the EL-photons are emitted within a small portion of the bubble, at the vicinity of the bottom of a hole [9], conditions may be found to push the interface deeper into the high-field region within a hole, thus enhancing the EL-photon yield. To this regard, an optimization of the electrode geometry allowing for an equilibrium configuration of the meniscus located at the hole center could be investigated (e.g. SC-GEM, [9]).

## Acknowledgments

We would like to thank our WIS colleagues: Dr. M. L. Rappaport (Weizmann Institute of Science - WIS) for all the precious advice with the LAr cryogenic system, Dr. E. Shimoni (WIS) for the assistance with the carbon-film evaporation and Y. Asher (WIS) for the technical support. Special thanks to Martin Kushner Schnur for supporting this research.

This work is supported in part by Sir Charles Clore Prize, by the Nella and Leon Benoziyo Center for High Energy Physics and by the Pazy Foundation.

# A Appendix

## A.1 Hydrostatic model for the position of the meniscus

We assess the admissible positions of a bubble meniscus within the THGEM hole using some approximations and a graphical solution approach.

We refer to the configurations shown schematically in Fig. 3. We identify univocally the menisci by the value of $z_t$ of the center of the cap on the symmetry axis, where the origin of $z$ is at the bottom face of the support. We assume that for every possible position the meniscus is a spherical cap, attaching to the solid boundaries with the prescribed contact angle $\theta$. In doing this we implicitly assume that the curvature $\kappa = 1/r$ of the meniscus is, for a given position, constant, and we neglect the hydrostatic pressure difference between the top and the contact line of the meniscus. In first approximation this difference is, in the relevant cases, of the order of or smaller than $(\rho_l - \rho_g)gR$, to be compared with $(\rho_l - \rho_g)gH$, and thus minor.

Due to convenience, we solve Eq. 2.4 graphically by plotting separately $p_b = p_1 - p_2$ and $p_L$, see Fig. 13. The intersections between the two curves give the sought solutions. Both pressure terms can be expressed as functions of the top bubble height $z_t$. The pressure difference between the liquid at the top of the bubble and the gas inside is simply $p_b = (\rho_g - \rho_l)g(z_t + H)$; the term $p_L(z_t)$ has a more complex dependence as a function of its location and concavity imposed by the contact angle constraint. To model the latter, we distinguish the three cases:

I. Meniscus attached to the upper face: for $z_t > z_4 = h + \frac{1-\cos\theta}{\sin\theta}(R+s)$, the meniscus has a negative curvature $\kappa = \frac{\cos\theta - 1}{z_t - h}$. This corresponds to $p_L(z) = 2\gamma\frac{\cos\theta - 1}{z_t - h}$, which is always directed downwards. Its lowest negative value is $p_L(z_4) = -2\gamma\frac{\sin\theta}{R+s}$.

II. Meniscus in the hole: at any height $z_2 = s + \frac{1-\sin\theta}{\cos\theta}R < z_t < z_3 = h - s + \frac{1-\sin\theta}{\cos\theta}R$ the meniscus attains a curvature $\kappa = -\frac{\cos\theta}{R}$, which accounts for $p_L = -2\gamma\frac{\cos\theta}{R}$, independent on $z_t$. This pressure is directed downwards for $\theta < 90°$ and upwards for $\theta > 90°$.

III. Meniscus attached to the lower face: for $z_t < z_1 = \frac{\cos\theta - 1}{\sin\theta}(R+s) < 0$, the meniscus has positive curvature $\kappa = \frac{1-\cos\theta}{z_t}$, corresponding to an upward pressure $p_L(z_t) = 2\gamma\frac{\cos\theta - 1}{z_t}$. The highest value of $p_L$ in this branch is obtained by the smallest spherical cap which attaches out of the rounded edge, that is $p_L(z_1) = 2\gamma\frac{\sin\theta}{R+s}$.

The relevant relations are obtained considering the circular arcs in Fig. 12, which meet the solid wall with the prescribed angle $\theta$. To account for the transitions between the cases, we assume that the edge of the hole is smooth and defined through a small radius of curvature $s \ll R$, see Fig. 12. The actual value of $s$ and the more irregular shape of a real hole are out of scope to this discussion. The smoothness of the edge is further important because it allows the attachment of a hemispherical meniscus of radius $\simeq R$ on the top side for $\theta < 90°$ (bottom for $\theta > 90°$), which corresponds to the maximal attainable curvature $\kappa = 1/R$, which is generally higher than the curvature in branch I (for $\theta < 90°$) or III (for $\theta > 90°$). This corresponds to a maximal containment pressure $p_L = -2\gamma/R$ in the first case (Fig. 12(a)), and to a maximal (unstable) contrasting pressure $p_L = 2\gamma/R$ in the other (Fig. 12(f)).

Summing up, a realization of the curves $p_L(z_t)$ and $p_b(z_t)$ is plotted in Fig. 13. The curves depend on all the parameters of the problem, and can accordingly intersect in 1 or 3 points, or even 5

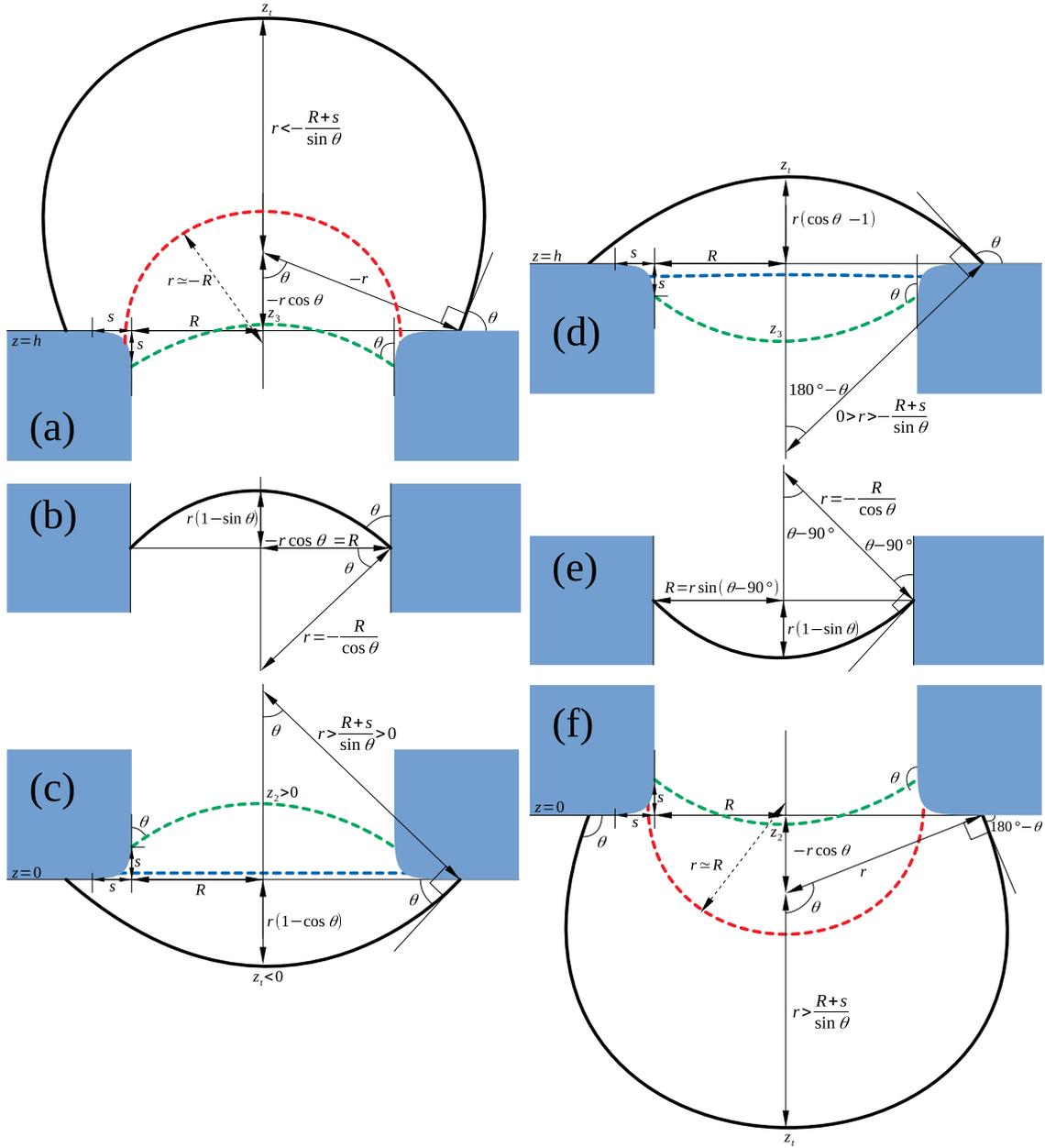

**Figure 12:** *Geometrical relations for the relevant meniscus positions: a)...c) for $\theta < 90°$, d)...f) for $\theta > 90°$, at the top, middle and bottom of the hole respectively. The green lines depict the topmost and lowermost acceptable menisci attached to the hole; the red lines show the maximally curved menisci attached to the smoothed edge, and the blue lines admissible flat menisci.*

– 20 –

for $\theta < 90°$. Moreover, an intersection corresponds to a stable equilibrium solution if $\frac{d(p_L - p_b)}{dz} < 0$, because a perturbation of the meniscus height $z$ would translate in a resulting restoring pressure, whereas the solution is unstable for $\frac{d(p_L - p_b)}{dz} > 0$. The graphical solution sketched in Fig. 13 shows that stable solutions, if they exist, fall within the transition zones I–II and II–III. While their existence is granted, their location is entirely dependent on the details of the edge smoothing. We thus expect that they are poorly defined in a real device, where the edges of the drilled holes are unavoidably irregular and different one from the other.

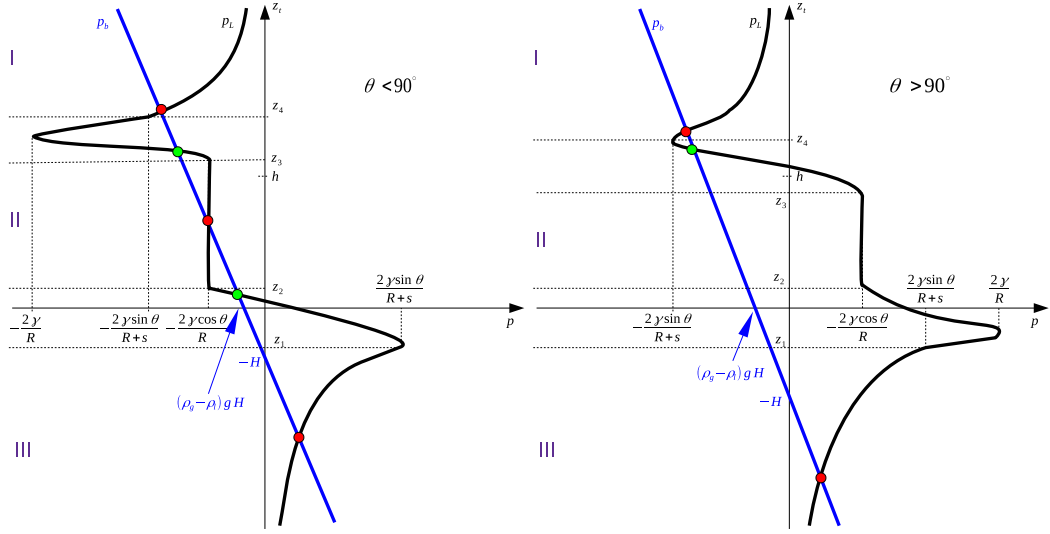

**Figure 13:** *Sample graphical solutions for $\theta < 90°$ (left) and for $\theta > 90°$ (right). The solid black curve and the blue line represent $p_L(z_t)$ and $p_b(z_t)$ respectively. Green dots represent stable solutions and red dots unstable ones. An interactive version of this figure, which allows to vary all the problem parameters, is available online as GeoGebra activity file,[33].*

In the ideal smooth edge case, some analytical conditions for the existence of stable solutions could be derived, but would lead to a complex set of inequalities, involving all the parameters of the problem and a transcendental dependence on $\theta$. We do not list here the expressions for all possible cases. As an example, however, for $\theta < 90°$ the condition for a stable meniscus pinned at the bottom side can be written as

$$(\rho_g - \rho_l)g(z_2 + H) > -2\gamma \frac{\cos\theta}{R}, \tag{A.1}$$

which translates into a condition for the bubble thickness H:

$$0 < H < 2\gamma \frac{\cos\theta}{R(\rho_l - \rho_g)g} - s - \frac{1 - \sin\theta}{\cos\theta} R \tag{A.2}$$

By the same token, a stable meniscus pinned at the top side would exist, for $\theta < 90°$, if

$$(\rho_g - \rho_l)(z_3 + H)g < -2\gamma \frac{\cos\theta}{R} \tag{A.3}$$

and

$$(\rho_g - \rho_l)(z_4 + H)g > -2\frac{\gamma}{R} \tag{A.4}$$

i.e.

$$2\frac{\gamma\cos\theta}{R(\rho_l-\rho_g)g} - h + s - \frac{1-\sin\theta}{\cos\theta}R < H < 2\frac{\gamma}{R(\rho_l-\rho_g)g} - h - \frac{1-\cos\theta}{\sin\theta}(R+s) \quad (A.5)$$

Using the material parameters of argon in the experimental conditions ($\gamma = 11.83\,\text{mN/m}$, $\rho_l - \rho_g = 1393\,\text{kg/m}^3$), $\theta = 48°$ as measured for Ar/FR4, and the typical dimensions of THGEM ($R = 150\,\mu\text{m}, h = 400\,\mu\text{m}, s \to 0$) the first condition (Eq.A.2) evaluates to $0\,\text{mm} < H < 7.7\,\text{mm}$, and the second (Eq.A.5) to $7.3\,\text{mm} < H < 11.1\,\text{mm}$.

## A.2 Methodology of the simulations

Simulations were performed using the microfluidics module of the commercial package COMSOL™(v. 5.4). To solve the two-phase flow problem, the Arbitrary Lagrangian-Eulerian grid (ALE) method was adopted. In this framework the computation is done in terms of Eulerian variables defined on a mesh whose deformation was somewhat arbitrary, while preserving faithfully the moving interface between the fluid and the gas domain. This is a sharp interface method; it was found that the other methods provided by COMSOL for two phase flow which make use diffuse interfaces, i.e. Level-Set or Phase-Field, suffer from known drawbacks which degrade the solution quality for the purposes of our investigation. Notably, diffuse interface methods are affected by spurious currents, which hinder the computation of quasi-static, capillarity-dominated flows, and require a high resolution in order to resolve fluid interfaces, which translates in prohibitive computational times. In contrast, ALE cannot not cope with large domain deformations, nor with domain topology changes, such as those taking place for systems of coalescing or bursting bubbles. Therefore we limited this investigation to the relaxation of the bubble meniscus in vicinity of the THGEM hole, which generally entailed small displacements.

For computational efficiency, a single bore THGEM was simulated using a 2D axysimmetric minimal domain, as depicted in Fig. 14. The coupled system of the Laplace equation for the electric potential $V$ and the Navier-Stokes equation for fluid velocity $u$ and pressure $p$ was solved. The material properties of argon at the operating conditions of Sec.3 were used. For the fluid part, open flow pressure boundary conditions were assigned to the top and bottom boundaries. A prescribed static angle of 48°, corresponding to the value measured on FR4, was assigned to the meniscus contact on the hole wall. For the electrostatic part, boundary conditions were assigned in order to emulate typical fields employed in THGEM detectors. Potentials were assigned at the top boundary, at the faces of the upper and lower electrode, and at the bottom boundary. The main control parameter was the voltage difference $V_0$ between the upper and the lower electrode. The top and bottom boundary potentials were assigned relatively to them, so that the mean electric field was $100\,\text{V/cm}$ above the THGEM and $250\,\text{V/cm}$ below it. These are typical values for the fields used in detectors, which drive the migration of electrons produced above the THGEM, and their actual value was of little importance in the present study. The electric force (2.1) on the interface between gas and liquid was implemented, in COMSOL parlancy, as a weak contribution on the fluid-fluid boundary, using test functions. The focus of this investigation is the effect of the electric field on the meniscus position, and hence the schematic boundary conditions imposed on the reduced domain may be somewhat inadequate for a real detector with a 3D structure. Complete simulations with the full geometry may eventually be done in the future; the setup of fields used here allows us already to shed light on the basic electro-hydrodynamics.

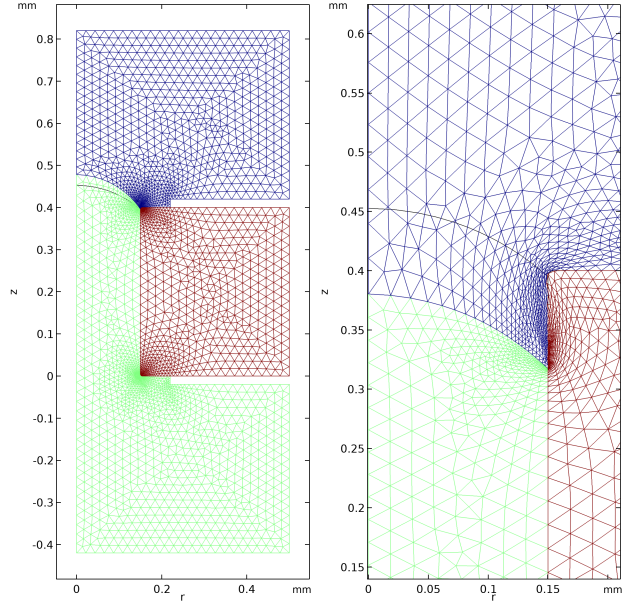

**Figure 14:** *Computational mesh for the case of meniscus at the top of the hole. Edges in blue, green and red refer to the liquid, gas and FR4 domain respectively. The black line is the initial meniscus position, in this case chosen arbitrarily as the topmost meniscus in the hole with contact angle $\theta$, corresponding to the hydrostatic solution for $H = 7.7$ mm. The plot on the left shows the mesh at the final time, for $H = 10$ mm and $V = 1200$ V. The plot on the right shows a zoom of the deformed mesh for the non-converging simulation with $H = 10$ mm and $V = 2800$ V.*

Since the sought equilibrium is unknown, the simulations were primed with the interface initially close to the expected equilibrium position, and the system was allowed to evolve dynamically. Control parameters of the simulation include the electrode voltages, the boundary pressures (equivalent to the bubble height $H$) and the contact angle $\theta$. The system was then integrated till $t = 100$ ms physical time. This simulation protocol was applied to bubble menisci placed either by the upper or by the lower face of the THGEM. This procedure was empirically seen as viable to let the system relax to the equilibrium, after a few damped meniscus oscillations. The lack of relaxation of the meniscus within this time (usually accompanied by an early numerical divergence, or by nonphysical results) was interpreted as an absence of a stable equilibrium for the meniscus for the probed parameter values. Fig. 14 exemplifies the well-behaved computational mesh produced at final time by a converging simulation, vs. the much too deformed mesh attained at an early time by a non-converging case, in which the meniscus is pushed downwards instead of settling at the top of the hole. This classification is certainly empirical, of relative value, and somehow dependent on the actual initial condition chosen, but it is still a criterion. Repeating the procedure for different values of the control parameters it was possible to discriminate intervals of existence of the equilibrium, like shown in Fig. 15 for varying $H$ and $V_0$.

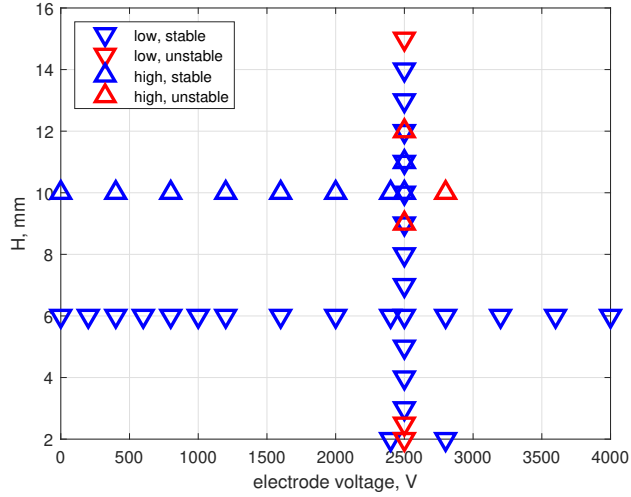

**Figure 15:** *Simulations converging to a stable equilibrium point or not converging, for various values of the bubble depth H and the potential $V_0$ on the electrodes. Upwards pointing triangles refer to simulations with the meniscus pinned at the top of the hole, and downwards pointing triangles at the bottom.*